\documentclass[12pt]{article}
\usepackage{amsmath,amssymb}

\usepackage{graphicx}
\usepackage{epstopdf}
\usepackage{wrapfig}

\usepackage{hyperref}
\usepackage{cite}

\def\beq{\begin{eqnarray}}
\def\eeq{\end{eqnarray}}

\newcommand{\be}{\begin{equation}}
\newcommand{\ee}{\end{equation}}
\newcommand{\bea}{\begin{eqnarray}}
\newcommand{\eea}{\end{eqnarray}}
\newcommand{\bg}{\begin{gather}}
\newcommand{\eg}{\end{gather}}
\newcommand{\bseq}{\begin{subequations}}
\newcommand{\eseq}{\end{subequations}}

\renewcommand{\ln}{\mathop{\rm ln}\nolimits}

\def\be{\begin{eqnarray}}
\def\ee{\end{eqnarray}}
\def\lb{\label}

\usepackage[font=small,labelfont=bf]{caption}

\begin{document}

\title{\textbf{Cosmological singularity,
conformal \\
anomaly  and symmetric polynomials}}
\vspace{1cm}
\author{ \textbf{Sergey N. Solodukhin  }} 

  \date{}
\maketitle
\begin{center}
    \emph{Institut Denis Poisson UMR-CNRS 7013,
  Universit\'e de Tours,}\\
  \emph{Parc de Grandmont, 37200 Tours, France} \\
\end{center}

{\vspace{-11cm}
\begin{flushright}
\end{flushright}
\vspace{11cm}
}



\begin{abstract}
\noindent {
We consider a spacetime singularity at $t = 0$ arising in a Kasner-type metric that solves the gravitational equations modified by quantum effects of a conformal field theory (CFT).
The resulting constraints can be solved efficiently when expressed in terms of symmetric polynomials.
Focusing first on the trace part of the modified gravitational equation, we determine the corresponding solution surfaces in the Kasner parameter space. The geometry of these surfaces depends sensitively on the ratio
$\eta = A/C$,
the quotient of the conformal charges characterizing the underlying CFT.
We then fully integrate the conformal anomaly near the singularity for a generic Kasner-type metric and obtain the corresponding stress-energy tensor. Its components are expressed in terms of three symmetric polynomials (of degrees $2$, $3$ and $4$) and depend on seven arbitrary constants, which may be interpreted as parameterizing different choices of the quantum state at the singularity. By imposing a set of constraints  we reduce this parameter space to a single free constant.
Subsequently, we solve, at leading order near the singularity, the modified gravitational equations. Among the admissible solutions, we identify, in particular, those that develop a curvature singularity while remaining geodesically complete.
}
\end{abstract}


\vskip 1 cm
\noindent
\noindent 
\noindent


\pagebreak

\newpage

\section{Introduction}
\setcounter{equation}0


The nature of spacetime singularities remains one of the most important open problems in gravitational physics. In the cosmological context, it is closely linked to the fundamental question of the origin of the observable Universe. Within classical general relativity, substantial progress has been made in identifying the general conditions under which singularities inevitably form. The classical theorems of Penrose and Hawking \cite{Penrose:1964wq} established that, under broad and physically well-motivated assumptions, the evolution of spacetime inevitably leads to regions in which curvature becomes unbounded and geodesics cannot be extended. In cosmology, such regions typically appear in the past of the Universe's time evolution, making it essential to understand the asymptotic behaviour of spacetime as one approaches the singularity. 

A major step toward such an understanding was achieved in the seminal works of Belinsky, Khalatnikov and Lifshitz (BKL) \cite{BKL}, who proposed a detailed asymptotic description of the metric near a spacetime singularity. According to the BKL scenario, time derivatives dominate over spatial ones as the singularity is approached, reducing the Einstein equations to a system of ordinary differential equations. In this regime, the metric evolves through a sequence of anisotropic contraction phases, the Kasner epochs, each characterized by a set of Kasner exponents and separated by abrupt transitions or bounces.The resulting oscillatory, and in fact chaotic, sequence of epochs is one of the most striking manifestations of nonlinear behaviour in general relativity. Further progress in the study of this chaotic dynamics was made in \cite{Damour:2000wm}.

However, it has long been suspected that the classical behaviour near singularities may be radically altered once quantum effects are taken into account. Insight into possible modifications can be gained by studying semiclassical gravity coupled to quantum conformal field theories, where the conformal anomaly provides a useful tool for analysing the backreaction of quantum matter on spacetime. Early work in this direction includes \cite{Starobinsky:1980te}-\cite{SS}.

In this paper, we systematically analyze the backreacted semiclassical  geometry within the class of Kasner-type metrics. We identify the constraints on the Kasner parameters arising from the conformal anomaly. We then integrate the anomaly explicitly in this class of backgrounds, which allows us to extract the full set of tensorial constraints. These constraints can be solved in closed form in terms of symmetric polynomials of the Kasner parameters.

Our initial motivation was to search for situations in which quantum CFT effects could regularize, or even forbid, the formation of singularities. These expectations turned out to be overly optimistic (except in the isotropic case, already understood in the context of the Starobinsky model \cite{Starobinsky:1980te}). Nevertheless, within the space of admissible Kasner parameters, we do find regions in which curvature singularities still form but the spacetime remains geodesically complete. Such spacetimes appear to be free of any known pathologies. 
The existence of these solutions in semiclassical gravity indicates that quantum effects may significantly modify the classical picture of singularity formation.
\section{A class of Kasner type  metrics}
\setcounter{equation}0
\subsection{ Metric and curvature}

In the present paper, we study a general class of cosmological anisotropic metrics which, when extended backward in time, exhibit a cosmological singularity at $t=0$. Near the singularity, and to leading order, the metrics we consider take the Kasner form,
\be
ds^2 = -dt^2 + t^{2a} dx_1^2 + t^{2b} dx_2^2 + t^{2c} dx_3^2 \, ,
\label{1}
\ee
where the constant parameters $(a, b, c)$ characterize a particular spacetime within the family. We emphasize that (\ref{1}) captures only the leading behavior of the metric in the limit $t \to 0$; at later times, subleading corrections, omitted here, will modify this form.

For the metric (\ref{1}), the components of the Einstein tensor $G^\mu_\nu=R^\mu_\nu -\frac{1}{2}\delta^\mu_\nu R$ are
\be
&&G^0_0=\frac{1}{t^2}(ab+ac+bc)\, ,\nonumber \\
&&G^1_1=-\frac{1}{t^2}(b^2+c^2+bc-b-c) \, , \nonumber \\
&&G^2_2=-\frac{1}{t^2}(a^2+c^2+ac-a-c)\, , \nonumber \\
&&G^3_3=-\frac{1}{t^2}(a^2+b^2+ab-a-b)\, ,
\lb{2}
\ee
and the Ricci scalar is
\be
R = \frac{1}{t^2}\bigl(a^2 + b^2 + c^2 + ab + bc + ac - a - b - c\bigr)\, .
\label{3}
\ee
These expressions give the leading contributions near the singularity.

\bigskip

\subsection{The regular spacetimes}

For generic values of $(a,b,c)$, the curvature is singular at $t=0$. To determine which values of the parameters yield a regular spacetime, we examine the quadratic curvature invariant constructed from the components of the Riemann tensor,
\be
R_{\alpha\beta\mu\nu}R^{\alpha\beta\mu\nu}
= \frac{4}{t^4}\Bigl[a^2(a-1)^2 + b^2(b-1)^2 + c^2(c-1)^2
+ a^2 b^2 + c^2 a^2 + b^2 c^2 \Bigr]\, .
\label{4}
\ee
Since this expression is a sum of non-negative terms, regularity requires that the divergent contribution in (\ref{4}) vanish, which is possible only if each term in the sum vanishes individually. This leads to two types of solutions: either all parameters vanish, $a=b=c=0$, or one parameter equals $1$ while the other two vanish, e.g. $(1,0,0)$.
In both cases, the spacetime described by (\ref{1}) near $t=0$ is free of singularities. Although these two forms of the metric appear different, they are in fact locally isometric and represent the same isotropic spacetime in different coordinate systems.

\subsection{ Regularity conditions for  singular metrics}
 
For all other values of the parameters $(a,b,c)$, the spacetime described by (\ref{1}) contains a singularity at (t=0). However, the singularity may be {\it milder}, or ``less singular'', provided that certain additional regularity conditions are satisfied. In what follows, we examine several such conditions.

\medskip

\noindent{\bf 1. Regular volume density and regular volume.}
For the metric (\ref{1}), the four-volume density is
$\sqrt{-g} = t^{a+b+c}$,
which has a finite limit as $t \to 0$ provided that
\be
a + b + c \ge 0 \, .
\label{4-1}
\ee
The classical Kasner spacetime is an example of such a geometry, with a regular (in fact, vanishing) volume density. In the space of all admissible parameters $(a,b,c)$, the set of spacetimes with regular volume density occupies a certain region; its complement consists of metrics for which the volume density diverges as $t \rightarrow 0$.
Requiring that the integrated four-volume,
$
V(t) \sim \int dt \sqrt{-g},
$
be finite leads to the condition
\be
a + b + c > -1 \, .
\label{4-2}
\ee
If this inequality is not satisfied, then the four-volume between any $t = t_0 > 0$ and $t = 0$ becomes infinite.

\medskip

\noindent{\bf 2. Integrated curvature is finite.} 
Although the square of the Riemann tensor behaves as $R_{\alpha\beta\mu\nu}^2 \sim t^{-4}$ and is therefore divergent at $t=0$, its integral
$\int dt\, \sqrt{-g} R_{\alpha\beta\mu\nu}^{2}$
can nevertheless be finite, provided that
\be
a + b + c > 3 \, .
\label{4-3}
\ee
For $a + b + c = 3$, the integral diverges logarithmically, whereas for $a + b + c < 3$ it diverges by a power law.
Similarly, requiring regularity of the integrated Ricci scalar,
$\int dt\, \sqrt{-g} R$,
imposes the bound
\be
a + b + c > 1 \, .
\label{4-4}
\ee
At the lower limit of this inequality, $a+b+c=1$, the integral diverges only logarithmically, which is milder than the power-law divergence encountered when the condition (\ref{4-4}) is violated.

\medskip

\noindent{\bf 3. Geodesic completeness.} 
Even if a spacetime possesses a curvature singularity, it may still be geodesically complete, provided that any lightlike geodesic reaches the singularity only at an infinite value of the affine parameter $\lambda$. In this sense, the divergent curvature is ``infinitely far away'' along null geodesics: no observer can physically reach it (or, rather, receive signals from it) within finite proper time or affine parameter.

Consider a light ray propagating along the $x^1$ direction. The change in the affine parameter between a point at finite $t$ and the singularity at $t=0$ behaves as $\Delta \lambda\sim \int t^a dt$.
This integral diverges when  $a \le -1$. Extending this argument to null rays propagating along any spatial direction $x^i\, i=1,2,3$, we obtain the more general condition,
\be
a\leq -1\, , \  \  b\leq -1\, , \  \  c\leq -1\, .
\lb{4-5}
\ee

\medskip

\noindent{\bf Geodesically complete spacetimes with curvature singularities.}
This analysis shows that for certain negative values of the parameters, the spacetime near the singularity attains infinite volume and its lightlike geodesics become complete. Since spacetime singularities are typically characterized by geodesic incompleteness, the case in which all parameters $(a,b,c)$ are negative and satisfy (\ref{4-5}) is particularly appealing, as it appears to evade the usual pathological features.

\bigskip

\subsection{Symmetric polynomials}

The metric (\ref{1}) is invariant under permutations of the spatial coordinates $(x^1, x^2, x^3)$. Consequently, any curvature invariant (with no free indices) must remain symmetric under permutations of the parameters $(a, b, c)$ and can therefore be expressed in terms of symmetric polynomials. It is thus convenient to introduce a basis in the space of symmetric polynomials. In the case of three variables, the independent symmetric polynomials are
\be
e_1=a+b+c\, ,  \  \  e_2=ab+bc+ac\, , \  \   e_3=abc\, .
\lb{5}
\ee
Other symmetric polynomials of a fixed degree can be constructed from these. In  degree 2 there are two independent elementary symmetric polynomials: $e_1^2$ and $e_2$.
In  degree 3 there are three polynomials: $e_1^3$, $e_1 e_2$ and $e_3$. In  degree 4 there are four: $e_1^4$, $e_1^2e_2$, $e_2^2$, $e_1 e_3$.
Below we provide some useful identities:
\be
&&a^2+b^2+c^2=e_1^2-2e_2\, , \nonumber \\
&&a^3+b^3+c^3=e_1^3-3e_1e_2+3e_3\, , \nonumber \\
&&a^2(b+c)+b^2(a+c)+c^2(a+b)=e_1 e_2-3e_3\, , \nonumber \\
&&a^4+b^4+c^4=e_1^4-4e_1^2e_2+2e_2^2+4e_1e_3\, , \nonumber \\
&&a^3(b+c)+b^3(a+c)+c^3(a+b)=e_1^2e_2-e_1 e_3-2e_2^2\, , \nonumber \\
&&a^2b^2+a^2c^2+b^2c^2=e_2^2-2e_1e_3\, .
\lb{6}
\ee
It is often convenient, and in some cases particularly useful, to regard the symmetric polynomials $(e_1, e_2, e_3)$ as independent variables instead of the parameters $(a, b, c)$. The Jacobian of this change of variables is proportional to $(a-b)(b-c)(a-c)$, and is therefore non-vanishing whenever the three parameters $(a, b, c)$ are distinct.

\bigskip

\noindent{\bf Consistency  bounds on symmetric polynomials.} One should stress that, although a priori parameters $(a,b,c)$ may take any real (positive or negative) values, it is not the case for the symmetric polynomials
$(e_1,e_2, e_3)$.  These satisfy  a number of bounds based on various mathematical inequalities, see \cite{Cauchy}. One restriction comes from the relation
\be
(a-b)^2+(b-c)^2+(c-a)^2=2(e_1^2-3e_2)\, .
\lb{x}
\ee
This implies the following bound
\be
e_2\leq \frac{e_1^2}{3}\, ,
\lb{6-1}
\ee
that restricts the possible values of $e_1$ and $e_2$.  As follows from (\ref{x}) the bound (\ref{6-1})  is saturated if and only if all three parameters are equal, $a=b=c$.

On the other hand, the inequality of arithmetic and geometric means  (known as the AM-GM inequality) implies that for any non-negative $a, b, c$  the following inequality holds
\be 
e_3\leq \frac{e_1^3}{27}\, ,
\lb{6-2}
\ee
and the equality holds if and only if $a=b=c$.
The other known inequality for non-negative $a, b, c$  is
\be
e_3^2\leq \frac{e_2^3}{27}\, ,
\lb{6-3}
\ee
the equality holds if and only if $a=b=c$.

\bigskip

\subsection{Classical Kasner metric}

In the classical case \cite{Kasner:1921zz}  the parameters $(a, b, c)$ are those for which  the metric (\ref{1}) solves   the Einstein equations in vacuum, $G^\mu_\nu=0$. 
The equation $G^0_0=0$ implies that the corresponding symmetric polynomial $e_2=0$.  The equation $G^1_1$ explicitly separates coordinate $x_1$ from $x_2$ and $x_3$ and hence 
the corresponding equation is symmetric under permutations of $b$ and $c$ but not under permutations that involve  parameter $a$. 
Indeed, as is seen from  (\ref{2}),  equation $G^1_1=0$ implies that
(taking into account that $e_2=0$)
\be
(a-e_1)(e_1-1)=0\, .
\lb{7}
\ee
This equation has two possible solutions. The first is $a=e_1$ that implies that $b+c=0$ and from constraint $e_2=0$ one finds that $bc=0$, i.e. $b=0$ or $c=0$. The equations $G^2_2=G^3_3=0$ then take the form
\be
a(a-1)=0\, .
\lb{8}
\ee
So that $a=0$ or $a=1$. Thus we find two possible solutions: $(0,0,0)$ and $(1,0,0)$ modulo the permutations. These are regular spacetimes as we discussed above.

The other solution of (\ref{7}) is $e_1=1$. This implies that equations $G^2_2=0$ and $G^3_3=0$ are satisfied automatically.  Thus, this second type solution defines $(a, b, c)$ for which
\be
e_1=1\, {\rm and} \, \, e_2=0\, .
\lb{9}
\ee
Equivalently, using  (\ref{6}), these constraints take the usual form
\be
a+b+c=1\, \, {\rm and} \, \,  a^2+b^2+c^2=1\, .
\lb{10}
\ee
The standard analysis shows that (provided that $a<b<c$): $-\frac{1}{3}\leq a\leq 0$, $0\leq b\leq \frac{2}{3}$ , $\frac{2}{3}\leq c\leq 1$. 
The equations (\ref{10}) can be solved by introducing a parameter $u$:
\be
a=-\frac{u}{w(u)}\, , \  \   b=\frac{1+u}{w(u)}\, , \  \   c=\frac{u(1+u)}{w(u)}\, , \  \   w(u)=1+u+u^2\, , \,  u\geq 1
\lb{11}
\ee
Geometrically, in the space of parameters $(a,b,c)$, it is represented by a circle that  is the intersection of a sphere of unit radius ($a^2+b^2+c^2=1$) and a plane ($a+b+c=1$).
The entire circle corresponds to all possible  values of the parameter $u$. 

Alternatively,  one can use the symmetric polynomials $(e_1, e_2, e_3)$  as independent variables. As we have seen already the symmetric polynomials $e_1=1$ and $e_2=0$. The polynomial $e_3$ takes negative values in the interval $-\frac{4}{27}\leq e_3\leq 0$.
The lower bound  is reached for values $(a=-\frac{1}{3}, b=c=\frac{2}{3})$.
So that in the space of variables $(e_1, e_2, e_3)$  the solution is given by an interval $(e_1=1, e_2=0, e_3=v\, , \, -\frac{4}{27}\leq v\leq 0)$.

The classical Kasner metric is characterised by a logarithmically divergent integral  of the Ricci scalar curvature (see (\ref{4-4})).
The classical equations  admit two regular spacetimes: $(0,0,0)$  and  $(1,0,0)$, they describe Minkowski spacetime.
The  other values of $(a,b,c)$ correspond to anisotropic  solutions. All of them are singular.

\section{Conformal anomaly and backreaction}
\setcounter{equation}0

\subsection{General set-up}
In the presence of quantum matter the gravitational equations are modified. In general, the gravitational action is a sum of the classical part and the part due to the
quantum matter, 
\be
W_{\rm gr}=-\frac{1}{2\kappa}\int \sqrt{-g}R +W_Q\, ,
\lb{12}
\ee
where $\kappa=8\pi G$.
For free quantum fields $W_Q=-\frac{1}{2}\sum_s (-1)^s\det \ln \Box_{(s)}$, where $s$ is the spin and $\Box_{(s)}$ is the respective field operator. Generally, $W_Q$ is a rather complicated 
non-local functional of the background  metric. In the case when the quantum fields are conformal an important local quantity that characterises $W_Q$ is available: the conformal anomaly
${\cal A}=g^{\mu\nu}\frac{\delta W_Q}{\delta g^{\mu\nu}}$.  That is why in the present paper we  focus on the case of conformal quantum field theory.

Variation of the gravitational action with respect to the metric gives us the modified Einstein equations,
\be
G^\mu_\nu=\kappa (T_Q)^\mu_\nu\, , \  \   (T_Q)_{\mu\nu}=\frac{\delta W_Q}{\delta g^{\mu\nu}}\, .
\lb{13}
\ee
We stress that the background metric is not fixed but appears as a solution to this  equation. We call it the {\it backreacted metric} as it includes the backreaction of the quantum fields
on the background metric.  The quantum stress-energy tensor $T^Q_{\mu\nu}$ 
is covariantly conserved,
\be
\nabla^\mu T^Q_{\mu\nu}=0\, .
\lb{14}
\ee
For the conformal quantum matter the trace of $T^Q_{\mu\nu}$ (henceforth we drop the subscript $Q$) is known to be a local function of metric,
\be
g^{\mu\nu}T_{\mu\nu}=-\frac{1}{(4\pi)^2} (A\,  E_4-C\, W^2)\, ,
\lb{15}
\ee
where $W^2$ is the square of the Weyl tensor and $E_4$ is the Euler density,
\be
&&E_4= R_{\alpha\beta\mu\nu}R^{\alpha\beta\mu\nu}-4R_{\mu\nu}R^{\mu\nu}+R^2\, , \nonumber \\
&&W^2=R_{\alpha\beta\mu\nu}R^{\alpha\beta\mu\nu}-2R_{\mu\nu}R^{\mu\nu}+\frac{1}{3}R^2\, .
\lb{16}
\ee
 $A$ and $C$ are the conformal charges that characterize a given conformal field theory. In a unitary theory these charges are positive. In this paper, however, we allow $A$ and $C$ to take arbitrary values, thereby effectively including the possibility of non-unitary theories as well as higher-spin theories \cite{AT} that may become relevant in the vicinity of the singularity.

\subsection{Our strategy}
Our primary goal is to solve the semiclassical gravitational equations  (\ref{13}) for the class of Kasner-type metrics  (\ref{1}) and to determine the constraints on the parameters 
$(a,b,c)$. We achieve this goal in two steps.

\medskip

\noindent{\bf Step 1.}  As already mentioned, the quantum stress tensor $T_{\mu\nu}$ is generally unknown and is defined up to  a number of ambiguities. However, its trace is a known local function of the metric.
Consequently,  the trace of the gravitational equations (\ref{13}) is local and unambiguous,
\be
R=\frac{\kappa}{(4\pi)^2}(A\, E_4-C\, W^2)\, .
\lb{17}
\ee
In the present section we shall analyze this equation for an anisotropic metric whose near-singularity behaviour is given by (\ref{1}).  To leading order, the left-hand side of equation (\ref{17}) behaves as $1/t^2$ in the vicinity of the singularity, whereas the right-hand side behaves as $1/t^4$. Thus, to leading order in $1/t$ equation (\ref{17}) for the metric in the general form (\ref{1}) reduces to  the requirement that the most divergent term on the right-hand side of  (\ref{17}) vanishes. This conditions leads to what we call the Master equation, which will be considered in the next subsection.

Proceeding with  Step 1 has the advantage that it does not involve any ambiguities related to the integration of the continuity equation (\ref{14}) for stress tensor $T^Q_{\mu\nu}$. 
However, the set of conditions on parameters $(a,b,c)$ obtained in Step 1 defines a   rather broad surface of admissible values in the parameter space.  In order to identify a more restricted  subset of these conditions,
we must consider the remaining tensorial components in  equations (\ref{13}). This is carried out  in Step 2.

\bigskip

\noindent{\bf Step 2.}  We first solve explicitly the continuity equation (\ref{14}) for the general form of the Kasner type metric (\ref{1}), without imposing a priori any constraints on parameters $(a,b,c)$.  This  procedure introduces several constants that remain  undetermined. Fixing these constants amounts to selecting  a particular quantum state
 of the conformal field. 
This analysis is carried out in Sections 4 and 5, where we focus on the leading divergent terms behaving as $1/t^4$.
The resulting stress tensor $T_{\mu\nu}$ is then substituted into equations (\ref{13}). As in Step 1, the left-hand side of (\ref{13}) behaves as $1/t^2$ while the right-hand side
behaves as $1/t^4$. Therefore, solving equation (\ref{13}) requires that the terms  in $T^\mu_\nu$ diverging as $1/t^4$ vanish. This condition effectively reduces to the vanishing of three symmetric polynomials in the parameters $a,b,c$, as we show in Section 5. The  solution of the resulting constraints is considered in Section 6.

\subsection{Master equation}
With these explanations, we proceed to solve the trace equation (\ref{17}).
First, we compute $W^2$ and $E_4$ near the singularity. For a Kasner type metric (\ref{1})
one finds that
\be
&&E_4=\frac{8}{t^4}\, a\, b\, c\, (a+b+c-3) \, , \\
&&W^2=\frac{4}{3t^4}\left((a-1)^2(b-a)(c-a)+(b-1)^2(a-b)(c-b)+(c-1)^2(a-c)(b-c)\right)\, . \nonumber
\lb{18}
\ee
We observe that the left-hand side of equation (\ref{17}), which originates from the Ricci scalar, behaves near the singularity as 
$1/t^2$, while the right-hand side, arising from the conformal anomaly, scales as $1/t^4$. 
Since the right-hand side diverges more strongly, consistency of the modified gravitational equations requires that it vanishes. This condition imposes a constraint on the parameters 
$(a,b,c)$, given by what we refer to as the Master equation:

\be
&&{\cal M}(a,b,c)\equiv 6\eta \, a\, b\, c\, (a+b+c-3)\nonumber \\
&&-\left((a-1)^2(b-a)(c-a)+(b-1)^2(a-b)(c-b)+(c-1)^2(a-c)(b-c)\right)=0\, ,
\lb{19}
\ee

\medskip

\noindent where we defined $\eta=A/C$ and we assumed that $C$ is non-vanishing.  This equation is symmetric under permutations of   parameters $a, b$ and $c$.

The case in which the conformal charge 
$C$ vanishes is, in a certain sense, degenerate. The the Master equation then takes the form
$$
a\, b\, c\, (a+b+c-3)=0\, ,
$$
which corresponds to four distinct planes in the parameter space 
$(a,b,c)$: namely, $a=0$, $b=0$, $c=0$ and $a+b+c=3$.
 Each of these planes satisfies the Master equation independently, while leaving the remaining parameters unconstrained.

\subsection{Some particular cases}

We now consider some particular  cases in which the Master equation simplifies.

\bigskip

\noindent{\bf Isotropic case: $a=b=c$.}
When all three parameters coincide the Weyl squared vanishes, $W^2=0$, so that the anomaly is determined only by the Euler term:
\be
{\cal M}(a,a,a)=18\eta  a^3(a-1)=0\, .
\lb{20}
\ee
One solution is $a=0$. It is the point $(0,0,0)$ in the space of parameters.
The spacetime in this case is regular at $t=0$. The other solution is $a=1$. It is the point $(1,1,1)$ in the space of parameters. There is a singularity at $t=0$ in this case.

\begin{figure}
\includegraphics[scale=0.6]{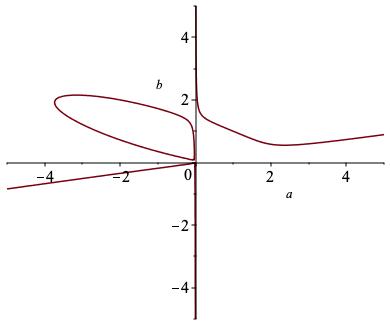}\hfill
\includegraphics[scale=0.6]{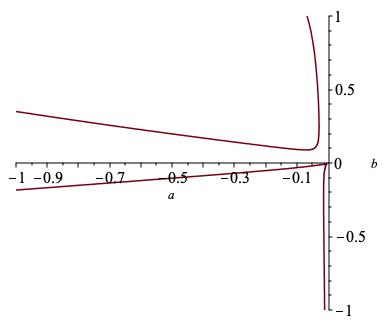}
\caption{\footnotesize Plot of curve  ${\cal M}(a,b)=0$ for value $\eta=3$ (left) and the zoomed-in view  of  the region of small negative values of $a$ (right).}
\label{Fig2-1}
\end{figure}

\begin{figure}
\begin{center}
\includegraphics[scale=0.7]{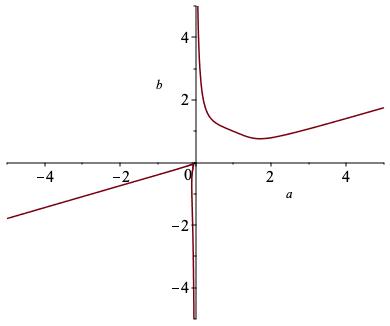}
\end{center}
\caption{\footnotesize Plot of curve  ${\cal M}(a,b)=0$ for value  $\eta=\frac{1}{3}$.}
\label{Fig2-2}
\end{figure}

\begin{figure}
\includegraphics[scale=0.6]{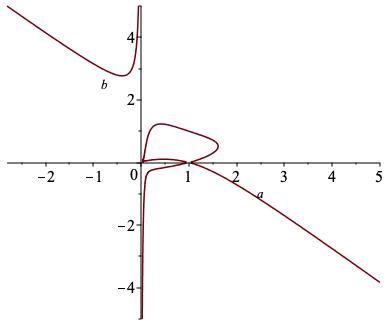}\hfill
\includegraphics[scale=0.5]{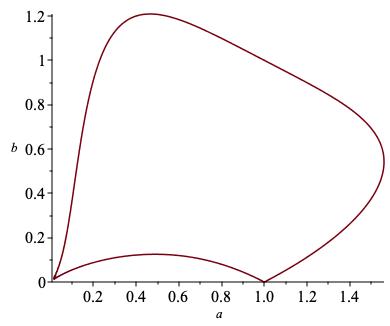}
\caption{\footnotesize Plot of curve  ${\cal M}(a,b)=0$ for value  $\eta=-\frac{1}{3}$ (left) and the zoomed-in view of the region of positive values of $(a,b)$ (right). }
\label{Fig2-3}
\end{figure}

\bigskip

\begin{figure}[htb]
 \centering
 \includegraphics[scale=0.6]{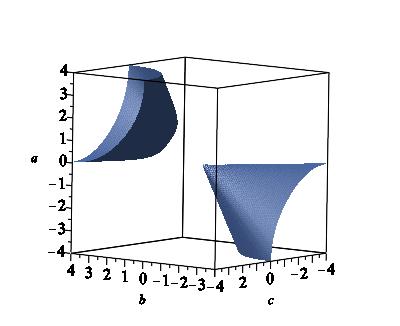}\hfill
  \includegraphics[scale=0.5]{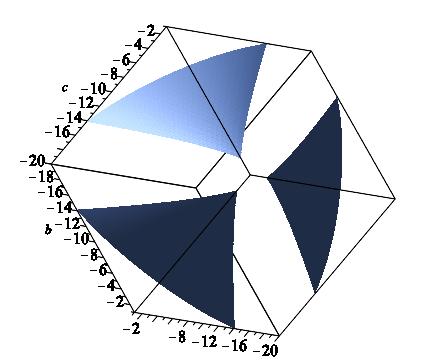}
 \caption{\footnotesize Surface ${\cal M}(a, b, c)=0$ for $\eta=A/C=1/2$.   Left: It contains of two open shells. One for positive values of  parameters $(a,b,c)$ and the other for all negative values. The conical point has coordinates $(0,0,0)$ in the parameter space.  Right: The three disjoint regions of the surface where each parameter is less than  $-1$, in the corresponding spacetime although the Riemann tensor is singular the geodesics remain complete.}
 \label{Fig3-1}
\end{figure}

\begin{figure}[htb]
 \centering
 \includegraphics[scale=0.6]{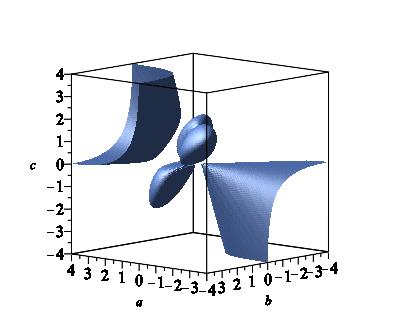}\hfill
  \includegraphics[scale=0.6]{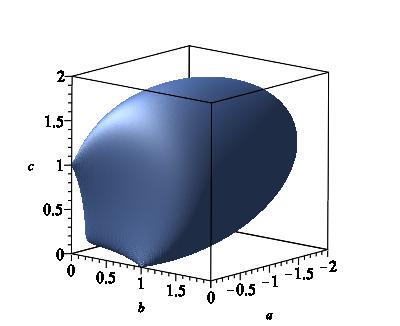}
 \caption{\footnotesize Left: Surface ${\cal M}(a, b, c)=0$ for $\eta=A/C=2$.   There appears three additional closed shells in which one of the parameters is negative and two others are positive.
 Right:  Zoomed-in view of such a shell for negative values of $a$ and positive values of $b$ and $c$.}
 \label{Fig3-2}
\end{figure}

\begin{figure}[htb]
 \centering
 \includegraphics[scale=0.6]{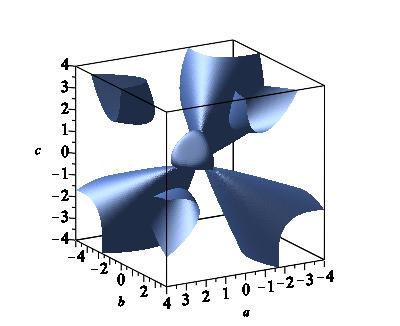}\hfill
  \includegraphics[scale=0.6]{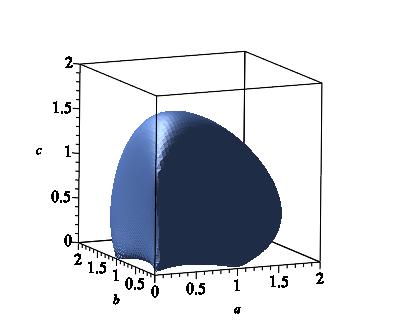}
 \caption{\footnotesize Surface ${\cal M}(a, b, c)=0$ for $\eta=A/C=-1$.  Left: Surface contains  an open shell in each region where one of parameters is negative and two others are positive and an open shell in each region where one of parameters is positive and two others are negative, and  a closed shell in the region where all parameters are positive. Right: Zoomed-in view of the region with the closed shell.}
 \label{Fig3-3}
\end{figure}

\noindent{\bf  Semi-isotropic case: $b=c\neq a$.} 
In this case $b=c\neq a$ and the Master equation becomes a function of two variables: $a$ and $b$,
\be
{\cal M}(a,b)=6\eta a b^2\, (a+2b-3)-(a-1)^2(b-a)^2=0\, .
\lb{20-1}
\ee
One obvious solution of this equation, valid for any value of $\eta$ is $(a=1, b=0)$. This corresponds to a regular spacetime $(1,0,0)$.
Analysis shows that this solution is isolated, i.e. it can not be continuously related to any other solution, for any positive value of $\eta$.

For negative $\eta$, however, it lies on a continuous curve that solves equation (\ref{20-1}). We present in Figures \ref{Fig2-1}-\ref{Fig2-3} the respective curves
that solve equation (\ref{20-1}) for various values of $\eta$. One sees that for $0<\eta< 1$ the curve has two open components: one lies in the domain  $a>0, b>0$
and the other in the domain $a<0, b<0$. 

For values $\eta> 1$ there a closed curve component appears 
in the domain $a<0, b>0$. This curve approaches quite closely the origin but does not in fact reaches it remaining always in the region $a<0, b>0$. 

For negative values $\eta<0$ the curve has three components: one open component in the domain $a<0, b>0$, another open component in the domain $a>0, b<0$ and a closed curve component
in the domain $a>0, b>0$ that passes by the points $(a=0, b=0)$ and $(a=1, b=0)$. These two points correspond to spacetimes without a singularity.

\subsection{General shape of the Master equation surfaces }
In a generic case a solution to the Master equation defines a two-dimensional surface in the space of parameters $(a,b,c)$. The shape of this surface crucially depends on the
parameter $\eta=A/C$ that is determined uniquely by the quantum conformal field theory in question. For a unitary CFT the parameter $\eta$ is positive and satisfies  certain
unitarity bounds that were studied in the literature. As we already explained this above in the present  study we do not restrict ourselves to only unitary theories
and thus prefer to explore all possible values of $\eta$ taking all possible (positive or negative) real numbers. 
In Figures \ref{Fig3-1}-\ref{Fig3-3} we present some typical situations with representative values of $\eta$.

 The analysis shows that as $\eta$ varies in the interval
$0< \eta<1$ the surface consists of two open shells.  One of them lies in the quadrant of positive values of parameters $(a,b,c)$ and is  bounded by plane $a+b+c=3$.
So that for all points in this branch one has that $a+b+c> 3$. The respective spacetime has regular integrated Riemann curvature, see (\ref{4-3}).

The second open shell lies in the region of negative values of all parameters $(a,b,c)$. It contains a conical point that has coordinates $(0,0,0)$.
It has three open infinite domains where the conditions (\ref{4-5}) to have  complete geodesics are satisfied. The respective spacetimes  have infinite 4-volume  near the singularity.

The entire surface is presented in Figure \ref{Fig3-1} for value $\eta=1/2$.  Note that the points where all parameters vanish or where one of parameters is $1$ and the others vanish are not shown since they lie outside the continuous surface and represent  isolated points in the parameters space.

For values  $\eta> 1$ the surface still has two open shells as in the case above. However, there also appear three closed identical shells each of them lying in a region where one of the parameters
is negative and the two others are positive. A surface of this type is shown in Figure \ref{Fig3-2}, where in particular  the closed shell  is shown in the region $(a<0, b>0, c>0)$.
The closed shell then passes by the points $(a=0, b=1, c=0)$ and $(a=0, b=0, c=1)$ that now are not isolated points and are continuously related to other points
in the closed shell component. Geometrically these three closed shells are identical due to the permutation symmetry in the parameter space $(a,b,c)$.

For negative values of $\eta$ the character of the surface changes again.  It contains an open shell in each quadrant where one of the parameter is negative and the two others are positive.
Then in each quadrant where one of the parameters is positive and the two others are negative there appears yet an open cone-like shell. Additionally, there appears a closed
surface component in the quadrant where all parameters are positive (shown in Figure \ref{Fig3-2} (right)). It passes by the points $(0,0,0)$, $(1,0,0)$, $(0,1,0)$ and $(0,0,1)$,
all four points representing the spacetime without singularities.

In summary, we find that the configurations, that are similar to  the classical Kasner solution, where one of the parameters is negative and the other two are positive, occur only for $\eta > 1$ and $\eta < 0$, and are absent for $0 < \eta < 1$. For $\eta > 1$, the corresponding points lie on a compact surface, so the Kasner parameters remain bounded. In contrast, for $\eta < 0$, the corresponding surface is non-compact, and the absolute values of the parameters are unbounded.

It should be noted that the two-dimensional surfaces that solve the Master equation represent only the trace part  of the modified Einstein equations, (\ref{17}).
The other tensorial components of the Einstein equations will produce further constraints on the parameters $(a,b,c)$ that may restrict
the possible solutions for parameters $(a,b,c)$ to a one-dimensional sub-set or even to some isolated points. This question will be further addressed
later in the paper.

\bigskip

\subsection{Solving the Master equation in terms of symmetric polynomials}

The Master equation (\ref{19}) can be re-written in terms of the symmetric polynomials $e_1, e_2$ and $e_3$,
\be
&&{\cal M}(e_1, e_2, e_3)\equiv 6(\eta-1)e_3(e_1-3)-h(e_1, e_2)=0\, ,\nonumber \\
&&h(e_1,e_2)=e_1^4-5e_1^2e_2+4e_2^2-2e_1^3+8e_1 e_2 +e_1^2-3 e_2\, .
\lb{21}
\ee

\bigskip

\noindent{\bf The case of $\eta=1$ ($A=C$).} This is the case of a special conformal field theory that plays an important role in the modern developments.
The superconformal 4d  $SU(N)$ Yang-Mills theory is a CFT of this type. This theory has a dual holographic description in the framework of the
AdS/CFT correspondence.

The first term in the Master equation (\ref{21}) in this case vanishes. The symmetric polynomial $e_3$ then disappears from  the Master equation so that it may take any value that is allowed.
 The Master equation reduces to equation  $h(e_1, e_2)=0$ that is a quadratic algebraic equation for $e_2$
and it can be easily solved,
\be
e_2^\pm(e_1)=\frac{1}{8}(e_1-1)\left(5e_1-3\pm 3\sqrt{(e_1-3)(e_1-\frac{1}{3})}\right)\, .
\lb{22}
\ee
First of all we notice that there is an isolated solution
\be
e_1=1\, , \  \   e_2=0\, .
\lb{23}
\ee
This is precisely the classical Kasner solution. It comes out  here because for $A=C$ the Riemann tensor disappears in  the conformal anomaly that now contains only Ricci tensor.
So that the solutions when the Ricci tensor vanishes are also solutions of equation (\ref{17}) with the conformal anomaly. This is the case for the Kasner solution.
This solution is isolated from any other possible solution. This is easily seen from the fact that the point $(e_1=1, e_2=0)$ lies outside of the  continuous curves defined by (\ref{22}). 

The solution (\ref{22}) has two branches depending on the sign. The expression under the square root is non-negative for two intervals: $e_1\geq 3$ and $e_1\leq \frac{1}{3}$.
For $e_1=0$ in the positive sign branch $e^-_2=0$. This is the only point when $e_2$ vanishes.  These values ($e_1=e_2=0$)  imply that the parameters $(a,b, c)$ may take only the value
$(0,0,0)$.  This is a  spacetime without any singularity.

For all other admissible values of $e_1$ within the two allowed intervals, $e_2^\pm$ remains positive for both branches. In the interval $e_1 \le \tfrac{1}{3}$, the negative branch lies above the positive one, $e_2^- > e_2^+$. Conversely, in the second interval, $e_1 \ge 3$, the ordering is reversed, with the positive branch lying above the negative one, $e_2^+ > e_2^-$.

\begin{figure}
\begin{center}
\includegraphics[scale=0.5]{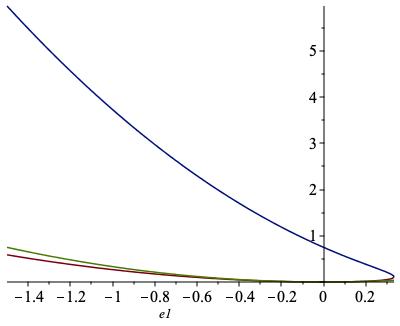}
\includegraphics[scale=0.5]{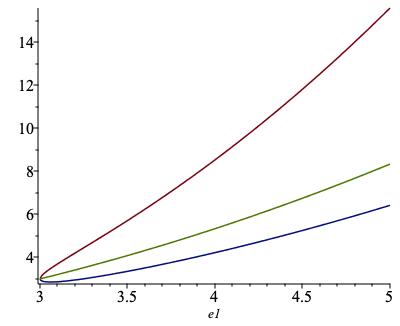}
\end{center}
\caption{\footnotesize Plot of functions $e^+_2(e_1)$ (black), $e^-_2(e_1)$ (bleu) and of  the bound $e_2^b(e_1)=\frac{e_1^3}{3}$  (green) for values $e_1\leq \frac{1}{3}$ (left) and $e_1\geq 3$ (right).} 
\label{Fig1}
\end{figure}

The bound (\ref{6-1}) imposes an additional constraint on possible values of  the symmetric polynomials. One finds that in the interval $e_1\geq 3$ the bound selects only the negative branch,
$e^-_2$.  In the interval $0 \leq e_1\leq \frac{1}{3}$  both branches violate the bound while in the interval $e_1\leq 0$ the positive sign branch satisfies the bound. This is shown  in Figure \ref{Fig1}.

Thus, we conclude that i) for values  $e_1\leq 0$ the bound selects the positive sign branch; ii) for values $0<e_1\leq \frac{1}{3}$  both branches violate the bound and thus there is no solution;
iii) for values $e_1\geq 3$ the bound selects the negative branch.

\bigskip

\noindent{\bf  The case  $\eta\neq 1$ ($A\neq C$).}  Consider now  a more general situation, when $\eta\neq 1$. 

\bigskip

\noindent i) First we consider  a special case when $e_1=3$. Then  the Master equation (\ref{21})  reduces to equation $h(e_2, e_1=3)=0$ with a single solution $e_2=3$.
Thus, in this case
\be
e_1=3\, , \  \   e_2=3\, .
\lb{23}
\ee
Note, that this solution saturates the bound (\ref{6-1}). It is possible only if all parameters $a, b, c$ are equal,
$a=b=c=1$.
We have already mentioned this solution above. It corresponds to a symmetric spacetime with a singularity at $t=0$.

\bigskip

\noindent ii) For $e_1\neq 3$,    the Master equation (\ref{21}) can be solved
as follows,

\medskip

\be
e_3=\frac{1}{6(\eta-1)} \frac{h(e_1, e_2)}{(e_1-3)}\, ,
\lb{24}
\ee

\medskip

\noindent where possible values of $e_1$ and $e_2$ are restricted by the bound (\ref{6-1}).
If values of $e_1$ lie in one of the intervals $e_1\geq 3$ or $e_1\leq \frac{1}{3}$  function $h(e_1, e_2) $ can be presented as
\be
h(e_1, e_2)=(e_2-e_2^+(e_1))(e_2-e_2^-(e_1))\, ,
\lb{25}
\ee
where $e^\pm_2(e_1)$ are defined in (\ref{22}). For $\frac{1}{3}\leq e_1\leq 3$, function  $h(e_1, e_2)$ does not vanish and takes positive values.

A particular solution is $e_1=1$, $e_2=0$ and $e_3=0$, this is point $(1,0,0)$ in the parameters space. It corresponds to a regular space-time.

\section{General form of anomalous stress tensor}

In the preceding sections, we analyzed the trace part of the modified gravitational equations that incorporate the conformal anomaly. The full semiclassical gravitational equations (\ref{13}), however, contain additional, tensorial, components that introduce further constraints. Geometrically, these constraints select a subspace of solutions within the surface ${\cal M}(a,b,c)=0$ considered earlier. To investigate their implications, one must first determine the general structure of the anomalous stress-energy tensor. In this section, we examine the complete set of equations for the stress-energy tensor
$T_{\mu\nu} = \bigl(T_{00}(t), T_{11}(t), T_{22}(t), T_{33}(t)\bigr)$
of the quantum CFT and provide a full integration of these equations.

The covariant conservation equation $\nabla^\mu T_{\mu\nu}=0$ in metric (\ref{1})   for $\nu=1,2,3$ is automatically satisfied (since $T_{0i}=0, \, i=1,2,3$) while for $\nu=0$
one has equation
\be
\partial_t T^0_0+\frac{(a+b+c)}{t}\, T^0_0-\frac{1}{t}(a\, T^1_1+b\, T^2_2+c\, T^3_3)=0\, .
\lb{26}
\ee
On the other hand, the trace anomaly equation (\ref{17}) leads to
\be
T^0_0+T^1_1+T^2_2+T^3_3=-\frac{C}{12\pi^2}\, {\cal M}(a,b,c)\, \frac{1}{t^4}\, ,
\lb{27}
\ee
where ${\cal M}(a,b,c)$ is defined in (\ref{19}).  Upon solving equation (\ref{26}) we represent the components of stress tensor as follows
\be
T^0_0(t)=-\frac{C}{12\pi^2}\frac{G(a,b,c)}{t^4}\, , \  \  T^i_i(t)=-\frac{C}{12\pi^2}\frac{F_i(a,b,c)}{t^4}\, , \  \   i=1,2,3
\lb{28}
\ee
Due to the permutation symmetry of the metric (\ref{1}), it is evident that  $G(a,b,c)$  is a symmetric function of the parameters 
$a$, $b$ and $c$.
 Additionally, the function  $F_1(a,b,c)$  is symmetric with respect to the permutation of 
$b$ and $c$.  The function  $F_2(a,b,c)$  is symmetric with respect to 
$a$ and $c$, while  $F_3(a,b,c)$ is symmetric with respect to  $a$ and $b$.

All three functions can be expressed in terms of a single function 
$F(a;b,c)$, which satisfies  $F(a;b,c)=F(a;c,b)$ as follows:
\be
F_1(a,b,c)=F(a;b,c)\, , \ \ F_2(a,b,c)=F(b; a,c)\, , \ \ F_3(a,b,c)=F(c; a,b)\, .
\lb{29}
\ee
Equations (\ref{26}), (\ref{27}) become algebraic equations for $G(a,b,c)$ and $F(a;b,c)$:
\be
G(a,b,c)+F(a;b,c)+F(b;a,c)+F(c;a,b)={\cal M}(a,b,c)\, 
\lb{30}
\ee
and
\be
&&F(a;b,c)(2a+b+c-4)+F(b;a,c)(2b+a+c-4)+F(c;a,b)(2c+a+b-4)\nonumber \\
&&=(a+b+c-4)\, {\cal M}(a,b,c)\, .
\lb{31}
\ee

\medskip

In the semiclassical Einstein equations (\ref{13}) modified by the contribution from a quantum conformal field theory (CFT), the left-hand side (due to the Ricci tensor) grows as 
$1/t^2$ for small  $t$, while the right-hand side, arising from the conformal anomaly, grows as 
$1/t^4$. The latter  therefore dominates   and must vanish in any consistent solution. To leading order, the Einstein equations thus reduce to a set of algebraic equations involving two functions of the parameters $(a,b,c)$:
\be
G(a,b,c)=0\, , \  \  F(a;b,c)=0\, , \ \  F(b;a,c)=0\, , \  \   F(c; a,b)=0\, .
\lb{32}
\ee
One may replace  equation $G(a,b,c)=0$ with  equation  ${\cal M}(a,b,c)=0$, i.e. that the anomaly vanishes to this order.

Our strategy now is to solve the algebraic equation (\ref{31}) and then identify the restrictions on parameters $a,b, c$ due to equations (\ref{32}).

\bigskip

\noindent{\bf Homogeneous solution.} 
We pause here for a brief remark. The differential equation (\ref{26}) also admits a homogeneous solution, which can always be added to the general solution of the non-homogeneous equation. This homogeneous solution corresponds to the case  $T^i_i=0\, , i=1,2,3$ in equation (\ref{26}).
 It takes the form $T^0_{0{hom}}\sim t^{-(a+b+c)}$.
The conformal anomaly does not contribute at this order unless 
$a+b+c=4$.  Therefore, we set $T^0_{0{hom}}=0$. 
The special case  $a+b+c=4$ requires separate treatment and will not be considered here.

\section{Kasner CFT}
\setcounter{equation}0
Both sides of equation (\ref{31}) represent a symmetric polynomial of degree 5 of three variables $a,b,c$. Any such polynomial can be decomposed over the symmetric basis
$e_1,e_2,e_3$, there are 16 terms in such a polynomial. Thus, equation (\ref{31}) is equivalent to a system of 16 linear equations. On the other hand, 
$F(a;b,c)$ is a polynomial of degree 4 symmetric in $b$ and $c$ and there may appear 22 terms  (and respective coefficients to be determined) in such a polynomial, see equation (\ref{1}) of appendix.
The naive counting, thus, suggests that there are 22-16=6  coefficients that remain undetermined. The direct analysis (see appendix A),  however, shows that there remain 7 undetermined coefficients, so that one of the 22 equations follows from the others.
In appendix A we solve equation (\ref{31}) and find the respective coefficients as functions of these 7 coefficients that remain undetermined (notice, however, that the choice of these coefficients is rather ambiguous).

With  a fourth-order polynomial $F(a; b,c)=F(a;c,b)$, one can define a third-order polynomial, $E(c;a,b)=E(c;b,a)$, and a second-order polynomial  $K(a; b,c)$ as follows
\be
&&F(a;b,c)-F(b;a,c)=(a-b)E(c;a,b)\, , \nonumber \\
&&E(c;a,b)-E(a;c,b)=(c-a)K(b; a,c)\, .
\lb{33}
\ee
For a generic polynomial $F(a;b,c)$ the second-order polynomial $K(b; a,c)$ does not have to be totally symmetric with respect to all its arguments $a,b,c$.
However, for  $F(a;b,c)$ that solves equation (\ref{31}) the respective second-order polynomial happens to be totally symmetric, $K(a;b,c)=K(a,b,c)$. 

One may define two more symmetric polynomials. The first is a third-degree polynomial,
\be
S(a,b,c)=E(a;b,c)+E(b;a,c)+E(c;a,b)\, ,
\lb{34}
\ee
and the second is a symmetric fourth-degree polynomial,
\be
H(a,b,c)=F(a;b,c)+F(b;a,c)+F(c;a,b)\, .
\lb{35}
\ee

The important point now is that  the polynomial $F(a;b,c)=F(a;c,b)$ can be expressed in term of these three totally symmetric polynomials, $K(a,b,c)$, $S(a,b,c)$ and $H(a,b,c)$,
\be
&&F(a;b,c)=\frac{1}{3}H(a,b,c)+\frac{1}{9}(2a-b-c)S(a,b,c)\nonumber \\
&&+\frac{1}{9}\left((a-b)(2c-a-b)+(a-c)(2b-a-c)\right)K(a,b,c)\, .
\lb{36}
\ee
This relation implies that, provided that $(a-b)(b-c)(c-a)$ is non-vanishing,  the equations $F(a;b,c)=F(b;a,c)=F(c;b,a)=0$ are equivalent to  the vanishing of  three symmetric polynomials:
\be
H(a,b,c)=0\, , \  \  S(a,b,c)=0\, ,  \  \   K(a,b,c)=0\, .
\lb{37}
\ee
As is shown in appendix for $F(a;b,c)$ that solve equation (\ref{31}) one finds that,
 \be
 K(a,b,c)=k_1(e_1^2-9)+\frac{1}{9}m_3(e_2-3)+k_3(e_1-3)\, . 
\lb{38}
\ee
and
 \be
 &&S(a,b,c)=m_1(e_1^3 -27)+m_2e_2(e_1-3)+m_3(e_3-e_2+2) \nonumber \\
 && +m_4(e_1^2-9) +m_5(e_1-3)+12(3-e_2)\, .
 \lb{388}
 \ee
The coefficients $k_1,k_3,m_1,m_2,m_3,m_4, m_5$ can be taken as the 7 coefficients that define all other coefficients in the polynomial $F(a;b,c)$.

For the symmetric polynomial $H(a,b,c)$ one finds (see appendix A) the following form
\be
&&H(a,b,c)=h_{21}(e_1^2-3e_2)+h_{31}(e_1^3-3e_1e_2)+h_{33}(e_3-\frac{1}{9}e_1e_2)\nonumber \\
&&+h_{41}(e_1^4-9 e_2^2)+h_{42}(e_1^2e_2-3e^2_2)+h_{44}(e_1e_3-\frac{1}{3}e_2^2)-2\eta(e_1e_2-\frac{3}{4}e^2_2)\, ,
\lb{39}
\ee
where $h_{21}$, $h_{31}$, $h_{33}$, $h_{41}$, $h_{42}$, $h_{44}$ are expressed in terms of the coefficients $k_1$, $k_3$, $m_1$, $m_2$, $m_3$, $m_4$, $m_5$
according to equation (\ref{a14}) in appendix A.

The relations (\ref{36}), (\ref{38}), (\ref{388}), (\ref{39}) completely solve the problem of finding the stress tensor with the conformal anomaly in a Kasner type metric (\ref{1})
and define the components of the CFT stress tensor according to equation (\ref{28}).

Examining equations (\ref{38}), (\ref{388}) and (\ref{39})  we can make the following observations. The second-order polynomial $K(a,b,c)$ does not contain any information about the conformal charges. The conformal charge $C$ (associated with the Weyl term in the anomaly) first appears in the third-order symmetric polynomial $S(a,b,c)$, while the conformal charge $A$ (associated with the Euler term) appears in the fourth-order polynomial $H(a,b,c)$. This pattern is reminiscent, albeit only loosely, of the familiar fact that the charge $C$ enters at the level of the two-point correlation function, whereas the charge $A$ enters at the level of the three-point function. We do not pursue this possible analogy further here.

\section{Solving the constraints}
\setcounter{equation}0
Thus, in the previous section we completely integrated the conformal anomaly for a Kasner type metric (\ref{1}), i.e. we have found the corresponding stress energy tensor that is covariantly conserved and its trace reproduces correctly the anomaly. Now, we want to solve the  complete gravitational equations considering the anomalous stress tensor as a  source for the gravitational field. These equations are tensorial. They include the vanishing of the anomaly, to leading oder in $t$, as one of the constraints.
As we have already explained in the previous sections,  the equations of semiclassical gravity, to leading order,  reduce to  the vanishing of three symmetric functions, $K(a,b,c)=0$, $S(a,b,c)=0$ and $H(a,b,c)=0$, where the functions are given be equations (\ref{38}), (\ref{388})
and (\ref{39}). They are defined up to 7 undetermined parameters,  $k_1$, $k_3$, $m_1$, $m_2$, $m_3$, $m_4$, $m_5$. If all these coefficients are non-vanishing then equation $K(a,b,c)=0$ is solved by expressing $e_2$ as a quadratic function of  $e_1$ which then  should be substituted into equation $S(a,b,c)=0$ that is solved by expressing $e_3$ as a cubic function of $e_1$.
All this should be then substituted into equation $H(a,b,c)=0$ that becomes a quartic equation for $e_1$. This equation may have at most 4 real solutions thus expressing $e_1$ as function of the 7 coefficients. Thus, in this case the solution of all three equations in the space of parameters $(a,b,c)$ is represented by four isolated points the position of which is determined by the values of the seven
coefficients.

This situation, however, is quite restrictive. Indeed, in the case of the classical Kasner metric the solution is a one-parameter family rather than a set of isolated points.
If we look at the classical gravitational equations (\ref{2}) we will find that in this case $F(a;b,c)=b^2+c^2+bc-(b+c)$ and hence one has that $E(a;b,c)=-(a+b+c-1)$.
This latter function is already symmetric in all three parameters and hence $K(a,b,c)=0$ identically. Then, one finds for the symmetric functions $S(a,b,c)=-3(a+b+c-1)$ and
$H(a,b,c)=2(a+b+c)^2-2(a+b+c)-3(ab+bc+ac)$. So that the equation $S(a,b,c)=0$ gives us condition $a+b+c=1$ and equation $H(a,b,c)=0$ gives us the second condition $ab+bc+ac=0$.
Both equations define a one-parameter family of solutions.

Although we do not fully understand the physical motivations for this prescription,  let us follow the classical case and assume that equation $K(a,b,c)=0$ holds identically and,  thus, does not
produce any constraints on parameters $a,b,c$. This can be achieved by the choice that the three coefficients in (\ref{38}) vanish, i.e.  $k_1=k_3=m_3=0$. In this case we have only two equations to define the possible values of parameters $a,b,c$: $S(a,b,c)=0$ and $H(a,b,c)=0$.  

 We may further restrict the number of non-vanishing coefficients. It is expected that the classical Kasner solution should be also a solution to the 
semiclassical gravitational equation if $\eta=A/C=1$. For a general choice of coefficients in (\ref{388}), (\ref{39}) this is not automatic. Indeed, imposing 
$e_1=1$ and $e_2=0$ in these equations we find that the corresponding solution exists provided a relation between the coefficients holds,
\be
m_6+13m_1+4m_4-18=0\, .
\lb{X}
\ee
This is a new constraint to be imposed on the coefficients. In order to further restrict the space of coefficients we further impose conditions that $m_1=m_4=0$ and $m_6=18$.
 With this choice of the coefficients, we arrive at the following expressions for the symmetric functions,
\be
&&S(a,b,c)\equiv e_2(m_2(e_1-3)-12)+18(e_1-1)=0\, ,  \lb{40} \\
&&H(a,b,c)\equiv \frac{9}{2}e_3(e_1-4)(\eta-1)-\frac{m_2}{12}e_2(e_1^2-3e_2)\nonumber \\
&&-\frac{3}{4}e_1^4+\frac{9}{4}e_1^3+\frac{15}{4}e_1^2e_2-\frac{35}{4}e_1 e_2-3e_2^2-\frac{3}{2}e_1^2+\frac{9}{2}e_2=0\, ,
\lb{41}
\ee
where we used equations (\ref{a14}). 

 As expected,  the system of equations $S(a,b,c)=0$ and $H(a,b,c)=0$ is equivalent to the Master equation ${\cal M}(a,b,c)=0$ (\ref{21}). Indeed, equation $S(a,b,c)=0$ can be used to
express $m_2$ in terms of $e_1$ and $e_2$ that can be substituted in equation (\ref{41}), one then obtains
\be
4(e_1-3)H(a,b,c)=3(e_1-4)\, {\cal M} (a,b,c)\, .
\lb{42}
\ee
So that $H(a,b,c)=0$ implies ${\cal M}(a,b,c)=0$ and vise versa (provided that  $e_1$ is not equal to $3$ or $4$).

For a given value of coefficient  $m_2$,  solving    equations (\ref{40}) and (\ref{41}) is straightforward.
The equation $S(a,b,c)=0$ can be easily solved to give (assuming that $m_2\neq -6$)
\be
e_2=-\frac{18(e_1-1)}{m_2(e_1-3)-12}\, \, .
\lb{42}
\ee
The expression (\ref{42})  then is  to be substituted in (\ref{41}) to give
\be
&&H(a,b,c)=-\frac{3(e_1-4)}{4(e_1m_2-3m_2-12)^2}\, Y(a,b,c)=0\, , \nonumber \\
&&Y(a,b,c)=-6e_3(m_2e_1-3m_2-12)^2(\eta-1)+(e_1-1)^2y(a,b,c)\, , \, \nonumber \\
&&y(a,b,c)= \left( m_2^2e_1^2(e_1-3)-6m_2(4e_1-3)(e_1-3)+72(2e_1-9)\right)\, .
\lb{43}
\ee
Provided that $e_1\neq 4$ we arrive at the equation,
\be
Y(a,b,c)=0\, .
\lb{44}
\ee

In the case when $\eta\neq 1$, equation (\ref{44})  
is a linear equation in terms of $e_3$ and its solution provides us with $e_3$ as a rational function of $e_1$,
\be
e_3=\frac{(e_1-1)^2 \, y(a,b,c)}{6(\eta-1)(m_2e_1-3m_2-12)^2}\, .
\lb{45}
\ee
Together with (\ref{42}) this gives us (for any given value of $m_2$) $e_2$ and $e_3$ as functions of $e_1$, the  possible value of which remains arbitrary subject to the consistency bounds that we
discussed earlier. We see that one obvious solution for any $\eta\neq 1$ is $e_1=1$, $e_2=0$, $e_3=0$.

In the case when $\eta=1$ 
this equation reduces to the equation $(e_1-1)^2y(a,b,c)=0$.  An obvious solution is $e_1=1$ (and $e_2=0$). This was our choice of the coefficients
in equations (\ref{388}) and (\ref{39}) to have this classical Kasner solution of the semiclassical equations when $\eta=1$.
For $e_1$ different from $1$ the equations reduce to $y(a,b,c)=0$. This is cubic equation.
Thus, it may have at most 3  real solutions for a given $m_2$.  For each value of $e_1$ the  corresponding value of $e_2$ then
is given by equation (\ref{42}). Value of $e_3$ remains arbitrary in this case. This is in full analogy with what we had in the classical case.
\begin{figure}[htb]
\centering
\includegraphics[width=.50\linewidth]{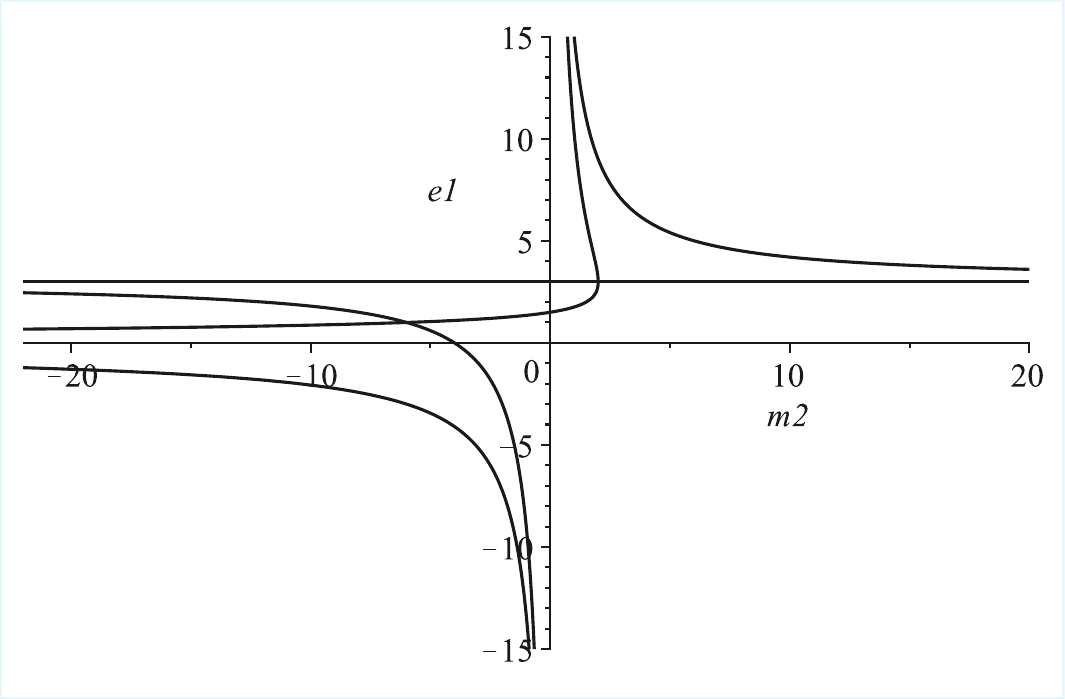}\hfill
\includegraphics[width=.50\linewidth]{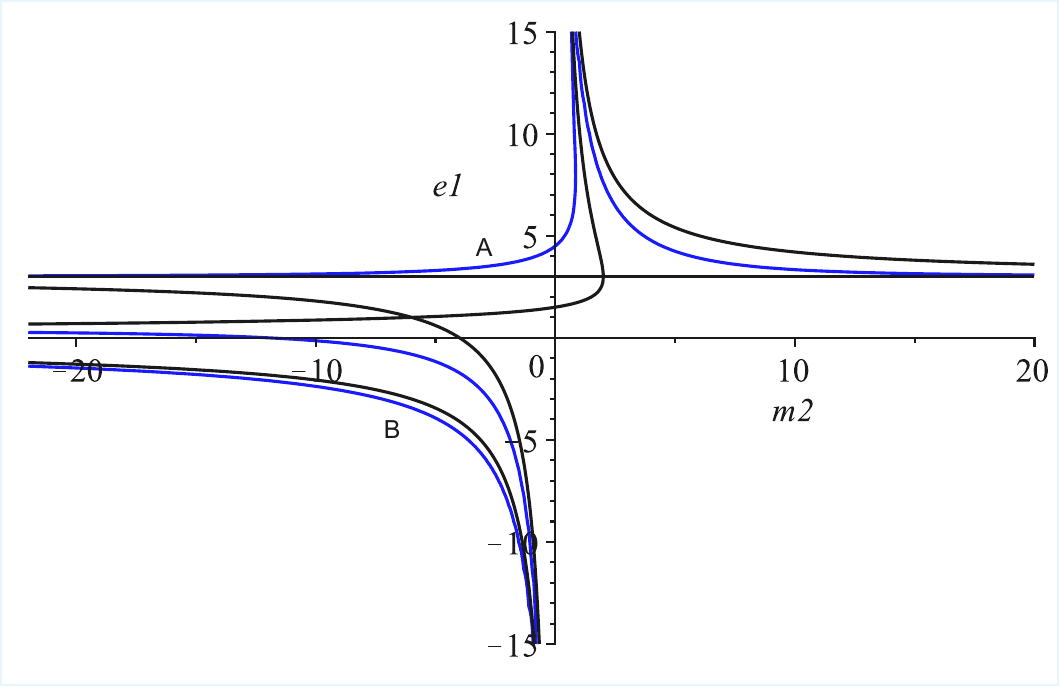}
\caption{\footnotesize 
Left: Plot of the zeros of the numerator and denominator vanishing of the rational function appearing on the left-hand side of (\ref{45}). The rational function is positive for large positive and large negative values of both $e_1$ and $m_2$, and it changes sign whenever a black contour is crossed.
Right: The plot additionally shows the four curves (blue) defined by $y(a,b,c)=0$. Among these, only the two curves labelled  ${\bf A}$ and ${\bf B}$ satisfy the consistency bound.
}
\label{Fig8}
\end{figure}

We recall that not any value of $e_2$ is possible since it should satisfy a consistency bound, $e_2\leq \frac{e_1^2}{3}$.  For $e_2$, given by equation (\ref{42}), we, thus, have the
following inequality,
\be
\frac{(e_1-3)(e_1^2 m_2-12e_1+18)}{(e_1m_2-3m_2-12)}\geq 0\, .
\lb{46}
\ee
This inequality should hold for any value of $\eta=A/C$. It imposes important constraints on the possible values of $e_1$. The corresponding allowed regions are shown in Figure \ref{Fig8}.  The inequality is satisfied for large values of $e_1$ and $m_2$ in the upper-right and lower-left corners of parameter plane and
its sign changes each time one crosses either black or the green line. Figure \ref{Fig8} also displays the four curves that solve equation $y(a,b,c)=0$, corresponding   to the case $\eta=1$. Among the four curves, only two of them (labeled A and B in the figure) satisfy the consistency bound.

\medskip

\noindent{\bf Singular but geodesically complete metrics as solutions.} It is instructive to examine more closely the limit of large negative values of  $e_1$.  This regime is reached for small negative values of $m_2$, since 
$e_1=\frac{12}{m_2}(1-\frac{1}{4}m_2-\frac{1}{12}m_2^2+O(m_2^3))$ as can be seen from equation $y(a,b,c)=0$. This is the blue curve labelled B in Figure \ref{Fig8}.
From  equation (\ref{42}) we find  that $e_2=\frac{e_1^2}{4}$ to the leading order, and the consistency bound is clearly satisfied.
As $e_3$ may take any permitted value we choose $e_3=se_1^3$, where $s>0$ is a parameter.Then the values of $a,b,c$ are the three roots of the cubic equation
$x^3-e_1x^2+e_2x-e_3=0$. This equation has three real roots $x_1<x_2<x_3$ provided $s<1/54$. All roots are negative:  $x_1<e_1/2$, $x_2<e_1/6$, $x_3<0$. For the third root
we have an estimate that $x_3\simeq 0,045 e_1$ provided $s=1/108$. 
 Hence, for large negative $e_1$  (small negative $m_2$) all three parameters $a,b,c$ that enter the Kasner
type metric (\ref{1}) take negative values that satisfy the condition (\ref{4-5}). The corresponding metric therefore possesses  a curvature singularity at $t=0$; however, this singularity lies at an infinite value of the affine parameter
for any null geodesic traced backward in time toward it.  In other words, any light emitted from the singularity can never reach any observer at a  later time. 
This result demonstrates that within the space of solutions to the semiclassical gravitational equations, there exists a domain of parameters in which the spacetime exhibits a curvature singularity while remaining geodesically complete.

This consideration can be extended to non-vanishing values of $\eta$ ($A\neq B$).  The solution (for a given value of $m_2$) in this case is fully specified by 
choosing a value of $e_1$ that satisfies condition (\ref{46}). Then values of  $e_2$ and $e_3$  are given by equations (\ref{42}) and (\ref{45}) respectively.
Provided that $m_2$ is negative and its value is sufficiently small we may choose 
\be
e_1=\frac{12}{m_2}-\lambda\, ,
\lb{47}
\ee
 where $\lambda$ is a free parameter to be further constrained. We find for the respective values of $e_2$ and $e_3$ to leading order, 
 \be
 e_2=\frac{3}{2(\lambda +3)} e_1^2\  \  {\rm and} \  \ e_3=\frac{(2\lambda+3)(\lambda-3)}{12(\lambda+3)^2(\eta-1)}e_1^3\, . 
 \lb{48}
 \ee
 The conditions  $e_2>0$ and $e_2\leq \frac{e_1^3}{3}$ are satisfied for $\lambda>3/2$. Then $e_3$ scales as $e_1^3$ for large negative values of $e_1$, the coefficient of proportionality is positive  if $\lambda>3$ (for $\eta>1$) or $3/2<\lambda < 3$ (for $\eta<1$). The bound (\ref{6-2}) imposes an extra constraint on the pre-factor in $e_3$ (\ref{48}).
Under these conditions   all three parameters $a,b,c$  take negative values that scale linearly with $e_1$ and hence satisfy the condition (\ref{4-5}). 
Whether all three parameters $a,b,c$ can simultaneously  be negative and less than $-1$
for other values of $m_2$ requires further investigation.

\section{Semiclassical solution in isotropic case: regular spacetime}
\setcounter{equation}0

In the previous section we mostly focused on the case when all three parameters $a,b,c$ take different values.  Now we want to discuss the isotropic case when all three parameters
have the same value, $a=b=c$. In section 3.3 we have already considered this case when we discussed the condition of vanishing of the anomaly, ${\cal M}(a,a,a)=18\eta a^3(a-1)$. 
There we concluded that this condition
selects two possible values $a=0$ ad $a=1$. The first is just regular spacetime that looks like Minkowski and the second has a curvature singularity. 

In the present section we would like to extend the discussion to include all tensorial components of the gravitational semiclassical equations. 

Let us return to the equations (\ref{30}) and (\ref{31}) for the functions $G(a; b,c)$ (related to component  $T^0_0$ of the anomalous stress tensor) and $F(a;b,c)$ (related to the component $T^a_a$).
As we discussed in sections 4, 5 and 6 solving these equations involves considerable  freedom. In the case when $a,b$ and $c$ take the same value this freedom disappears and the result is unique. Indeed, the equations (\ref{30}) and  (\ref{31}) become equations
\be
&&G(a)+3 F(a)=18\eta a^3(a-1) \, \nonumber \\
&&12F(a)(a-1)=(3a-4)18\eta a^3(a-1)\, ,
\lb{49}
\ee
where we use notations $G(a)=G(a;a,a)$ and $F(a)=F(a; a,a)$.  This has the unique solution,
\be
F(a)=\frac{3}{2}\eta (3a-4)a^3\, ,  \  \   \  G(a)=\frac{9}{2}\eta a^4\, .
\lb{50}
\ee
We stress that this solution is unambiguous: no free constants are involved.
The gravitational equations then reduce to  $F(a)=0$ and $G(a)=0$. We find that only the value $a=0$ remains as a valid solution while
$a=1$ appears to be spurious.  Thus,   only the isotropic spacetime, which is free of singularities, satisfies the complete  semiclassical equations.
This is, of course, a well-known feature of the Starobinsky model \cite{Starobinsky:1980te}. A detailed analysis of the isotropic case is presented in Appendix B, where we show that the absence of the singularity arises from a particular choice of quantum state. The still remaining freedom is in the
solution of the homogeneous differential equation that we systematically neglected in the paper. In the isotropic case adding this solution is equivalent to adding an isotropic radiation to the quantum state. This generically leads to appearance of the singularity. On the other hand, as was explained in \cite{Starobinsky:1980te}, the absence of such radiation
in the quantum state leads to a completely regular isotropic solution of gravitational equations. 

\section{Conclusions}
\setcounter{equation}0

In the present paper, we address the problem of cosmological singularities in semiclassical gravity modified by a quantum conformal field theory (CFT). For Kasner-type metrics, the leading-order behavior in the vicinity of the singularity reduces the problem to a set of algebraic constraints imposed on the Kasner parameters. We determine the complete asymptotic structure of the anomalous stress-energy tensor of the quantum CFT, an uncommon situation, as such analyses can be carried out explicitly only in a few very special cases. Integrating the conformal anomaly for Kasner-type metrics allows us to obtain the full set of constraints on the parameters, which can be solved conveniently when expressed in terms of symmetric polynomials.

The classical Kasner solution appears as a special case of our solutions when  $\eta =A/C =1$ ($A$ and $C$ are the conformal charges), which in particular includes the holographic CFT. In addition to this classical branch, we find a family of solutions that differ substantially from the classical case. Classically, the Kasner parameters are bounded and lie within a finite domain. In contrast, the solutions to the semiclassical equations are generically unbounded. Configurations in which one parameter is negative and the other two are positive, typical of the classical case, are not generic here and, in fact, do not appear when $0 < \eta < 1$. Such configurations arise  when $\eta > 1$ and $\eta<0$. 

Moreover, we identify open regions in the parameter space in which all three Kasner parameters are either negative or positive. This suggests that the asymptotic approach to the singularity may differ significantly from the usual BKL scenario. Determining its precise nature requires further investigation based on the results presented here.

Another interesting finding is the existence of solutions in which all three Kasner parameters are negative and less than $-1$. We provide explicit examples of such solutions. These solutions are notable because they represent a new type of resolution to the cosmological singularity problem. Although a curvature singularity at $t = 0$ still exists, the geodesics are complete, placing the singularity at infinite affine distance from any physical observer.  This scenario deserves a more detailed examination.

\bigskip
\section*{Acknowledgements} 
The author would like to thank the Isaac Newton Institute for Mathematical Sciences, Cambridge, for support and hospitality during the programme  {\it  Quantum field theory with boundaries, impurities, and defects}, where work on this paper was completed. This work was supported by EPSRC grant EP/Z000580/1.

\newpage
\appendix
\section{Function $F(a;b,c)$ for a Kasner type metric}
\setcounter{equation}0
\numberwithin{equation}{section}

In this appendix we solve equation (\ref{31}) and find function $F(a; b, c)$.  We consider generic variables $x, y, z$ and function 
$F(z; x,y)=F(z;y,x)$ symmetric in last two arguments and assume that it is a polynomial of degree 4 of  its arguments. Since it is symmetric in $x$ and $y$
it can be decomposed in powers of symmetric basic polynomials $(x+y)$ and $xy$. It can be thus represented as follows
\be
F(z; x, y)=\sum_{n=0}^4 F_{(n)}\, , \  \   F_{(n)}=\sum\limits_{\substack{(l)\\ k+p+q=n}} f_{nl}\, z^k (y+x)^p(yx)^q\, ,
\lb{a1}
\ee
where $f_{nl}$ are the coefficients to be determined, index $l$ changes from $1$ to  a maximum value that depends on number of independant 
polynomials at order $n$. More precisely we have that
\be
&&F_{(0)}=f_{0}\, , \  \   F_{(1)}=f_{11}(x+y)+f_{12} z\, , \  \  F_{(2)}=f_{21}(y+x)^2+f_{22} xy+f_{23} z^2\, ,  \\
&&F_{(3)}=f_{31}(y+x)^3+f_{32}(x+y)xy+f_{33}xyz+f_{34}z(x+y)^2+f_{35}z^2(x+y)+f_{36}z^3\, , \nonumber\\
&&F_{(4)}=f_{41}(x+y)^4+f_{42}(x+y)^2 xy+f_{43} (xy)^2+f_{44}z(x+y)^3+f_{45}z(x+y) xy \nonumber \\
&&+f_{46}z^2(x+y)^2+f_{47}z^2 xy+f_{48}z^3(x+y)+f_{49}z^4\, . \nonumber
\lb{a2}
\ee
This representation has to be substituted into equation (\ref{31}). This equation is symmetric in $a, b$ and $c$ and thus can be decomposed over  the basis $e_1$, $e_2$ and $e_3$ defined earlier in (\ref{5}).  This equation is a polynomial of degree 5 and thus has 16 independent terms.  That means that equation (\ref{31}) reduces to
16  different equations for the coefficients $f_{nl}$. In (\ref{a2}) there are 22 such coefficients. A naive counting would suggest that there remain 22-16=6  undetermined
coefficients. The direct calculation however shows that there are 7 such coefficients. It appears that one of the equations follows from the others.
There is an ambiguity which coefficients are to be taken as these 7 coefficients that define all other coefficients. Here is one possible choice to solve the system of linear equations that come from equation (\ref{31}):
\be
&&f_{0}=0\, , \  \  f_{11} =-9+24\eta+8f_{34}+2f_{24}+4f_{33}\, \\
&& f_{12} =18-48\eta-16f_{34}-4f_{24}-8f_{33}\nonumber\\
&&f_{21} = -108+12\eta-4f_{34}+56f_{48}-184f_{49}-f_{24}-8f_{45}+32f_{42} \nonumber \\
&&f_{22} = 425/2-12\eta+f_{24}+12f_{34}+2f_{33}-112f_{48}+368f_{49}+16f_{45}-64f_{42}\nonumber \\
&&f_{23} = 439/2-36\eta+4f_{34}+f_{24}-2f_{33}-112f_{48}+368f_{49}+16f_{45}-64f_{42}\nonumber\\
&&f_{31} = 57-12\eta-28f_{48}+94f_{49}+4f_{45}-16f_{42}-f_{33}\nonumber \\
&&f_{32} = -116+18\eta-3f_{34}+f_{33}+60f_{48}-196f_{49}-8f_{45}+32f_{42}\nonumber \\
&&f_{35} = -84+38f_{48}+21\eta +f_{34}-132f_{49}-6f_{45}+24f_{42}+2f_{33}\nonumber \\
&&f_{36} = -56+12\eta+28f_{48}-96f_{49}-4f_{45}+16f_{42}+f_{33}\nonumber\\
&&f_{41} = -1/2-f_{49}\, , \  \  f_{44}= 7/2+6f_{49}-f_{42}-2f_{48}\nonumber \\
&&f_{43} = -59/2+9/2\eta-47f_{49}+8f_{42}+15f_{48}-3f_{45}\nonumber \\
&&f_{46} = 19/2-3/2\eta+15f_{49}-3f_{42}-4f_{48}+f_{45} \nonumber \\
&&f_{47} = 6\eta-57/2-46f_{49}+8f_{42}+15f_{48}-3f_{45}\nonumber 
\lb{a3}
\ee
The seven coefficients $f_{34}$, $f_{24}$, $f_{33}$, $f_{42}$, $f_{45}$, $f_{48}$, $f_{49}$ remain undetermined by the equations.

From the 4th order polynomial $F(a;b,c)$ one may define  a 3rd order polynomial $E(c; a,b)=E(c;b,a)$ by  the relation
\be
 F(a; b, c)-F(b; a, c)=(a-b)E(c; a,b)
 \lb{a4}
 \ee
and then also a 2nd order polynomial  $K(b;a,c)$ defined as follows
\be
E(c;a,b)-E(a;c,b)=(c-a)K(b; a,c)\, .
\lb{a5}
\ee
It is an important fact that for $F(a;b,c)$ defined by (\ref{a1})-(\ref{a3})  the polynomial $K(b; a,c)$ is  symmetric in $(a,b,c)$ so that $K(b; a,c)=K(a,b,c)$.
Additionally, there is a constraint on the coefficients in the decomposition over a symmetric basis, the exact form then reads
\be
K(a,b,c)=k_1(a^2+b^2+c^2)+k_2(ab+ac+bc)+k_3(a+b+c)-3(k_1+k_2+k_3)\, .
\lb{a6}
\ee
For the coefficients one finds,
\be
&&k_1= 2f_{42}+2f_{48}-8f_{49}-5 \\
&&k_2 = 14f_{42}-4f_{45}+22f_{48}-72f_{49}+6\eta-45\nonumber\\
&&k_3 = -4f_{34}+16f_{42}-4f_{45}+32f_{48}-104f_{49}+6\eta-58 \nonumber
\lb{a7}
\ee
On the other hand, the symmetrisation of polynomials $E(a;b,c)$ in
 all three arguments produces a 3rd order symmetric polynomial
 \be
 S(a,b,c)=E(a;b,c)+E(b;a,c)+E(c;a,b)
 \lb{a8}
 \ee
 that can be decomposed over the basis $e_1$, $e_2$ and $e_3$. 
 There are in general 7 coefficients in this decomposition,
 \be
S(a,b,c)= e_1^3m_1+e_1e_2m_2+e_3m_3+ e_1^2m_4+e_2m_6+e_1m_5+m_7\, .
 \lb{a88}
 \ee
 The direct  calculation reveals some algebraic relations between the coefficients
 \be
 &&m_6+m_3+3m_2+12=0\, , \nonumber \\
 &&m_7+27m_1+3m_5+9m_4+2m_3-36=0\, .
 \lb{a89}
 \ee
 One can use these relations to express $m_6$ and $m_7$ in terms of other coefficients (this is of course an ambiguous choice). This leads to
 leads to the following form
 \be
 &&S(a,b,c)=m_1(e_1^3 -27)+m_2e_2(e_1-3)+m_3(e_3-e_2+2)\nonumber\\
 && +m_4(e_1^2-9) +m_5(e_1-3)+12(3-e_2)
\lb{a9}
\ee
with only five coefficients that are defined as
\be
&&m_1 = -4f_{48}+28f_{49}-4f_{42}+13 \\
&&m_2 = 43-18f_{48}+52f_{49}-14f_{42}+8f_{45}-12\eta \nonumber\\
&&m_3 = 90f_{42}-36f_{45}+162f_{48}-504f_{49}+54\eta-315\nonumber\\
&&m_4=272f_{48}-932f_{49}+132\eta+12f_{33}+8f_{34}+160f_{42}-40f_{45}-562\nonumber\\
&&m_5=-672f_{48}+2208f_{49}-216\eta+12f_{24}-12f_{33}+24f_{34}-384f_{42}+96f_{45}+1317\nonumber
\lb{a10}
\ee
in terms of the independent  coefficients $f_{nl}$.
 
In fact, it is convenient to choose $k_1$, $k_3$, $m_1$, $m_2$, $m_3$, $m_4$ and  $m_5$ as independent variables and express all  coefficients
$f_{nl}$ in terms of them,
\be
&&f_{24} = \frac{1}{4}m_1+2k_1+\frac{1}{12}m_4+\frac{1}{18}m_3+\frac{2}{3}k_3+\frac{1}{12}m_6\nonumber\\
&&f_{33} = \frac{1}{12}m_4+\frac{1}{4}m_1+\frac{7}{4}-\frac{3}{2}k_1-\frac{1}{3}m_2-\frac{5}{27}m_3-6\eta+\frac{1}{6}k_3  \nonumber\\
&&f_{34} = \frac{1}{6}m_2+\frac{1}{4}k_1+\frac{7}{108}m_3-\frac{1}{4}k_3 \nonumber\\
&&f_{42} = \frac{1}{12}m1+\frac{5}{4}+\frac{11}{12}k_1-\frac{1}{12}m_2-\frac{1}{54}m_3  \nonumber\\
&&f_{45} = \frac{1}{6}m_2+\frac{13}{12}k_1+\frac{1}{108}m_3+\frac{1}{6}m_1-1+\frac{3}{2}\eta   \nonumber\\
&&f_{48} = \frac{1}{4}m_1+\frac{1}{4}+\frac{1}{4}k_1+\frac{1}{12}m_2+\frac{1}{54}m_3  \nonumber\\
&&f_{49} = \frac{1}{12}m_1-\frac{1}{4}+\frac{1}{6}k_1
\lb{a11}
\ee
One also finds that
\be
k_2 = \frac{1}{9}m_3+2k_1\, .
\ee
Then, one has  that for the 2nd order symmetric polynomial,
\be
K(a,b,c)=k_1(e_1^2-9)+\frac{1}{9}m_3(e_2-3)+k_3(e_1-3)\, ,
\lb{a11}
\ee
if it is expressed in terms of the basis of symmetric polynomials.

\bigskip

The next symmetric polynomial, of degree 4, is defined as
\be
H(a,b,c)=F(a;b,c)+F(b;a,c)+F(c;a,b)\, .
\lb{a12}
\ee
It can be decomposed over the symmetric basis
\be
&&H(a,b,c)=h_0+h_1e_1+h_{21}e_1^2+h_{22}e_2+h_{31}e_1^3+h_{32}e_1e_2+h_{33}e_3\nonumber\\
&&+h_{41}e_1^4+h_{42}e_1^2e_2+h_{43}e^2_2+h_{44}e_1e_3\, .
\lb{a13}
\ee
The coefficients $h_{nl}$ then are expressed in terms of $k_1$, $k_3$, $m_1$, $m_2$, $m_3$, $m_4$ and  $m_5$ as follows
\be
&&h_0=0\, , \  \  h_1=0\, , \  \  h_{21} = -\frac{3}{4}m_1-\frac{1}{12}m_5-\frac{1}{4}m_4+\frac{1}{18}m_3\nonumber \\
&&h_{22} = \frac{1}{4}m_5-\frac{1}{6}m_3+\frac{9}{4}m_1+\frac{3}{4}m_4\nonumber \\
&&h_{31} = -\frac{1}{12}m_4-\frac{1}{2}k_1-\frac{1}{6}k_3-\frac{1}{4}m_1+\frac{9}{4}\nonumber \\
&&h_{32} = \frac{9}{4}k_1+\frac{1}{36}m_3+\frac{3}{4}k_3+\frac{3}{4}m_1+\frac{1}{4}m_4-\frac{35}{4}  \nonumber \\
&&h_{33} = -18\eta-\frac{27}{4}k_1-\frac{1}{4}m_3-\frac{9}{4}k_3+18 \nonumber \\
&&h_{41} = -\frac{1}{6}k_1-\frac{1}{12}m_1-\frac{3}{4}  \nonumber \\
&&h_{42} = \frac{3}{4}k_1-\frac{1}{54}m_3+\frac{1}{4}m_1-\frac{1}{12}m_2+\frac{15}{4} \nonumber \\
&&h_{43} = \frac{1}{12}m_3+\frac{1}{4}m_2-3 \nonumber \\
&&h_{44} = \frac{9}{2}\eta -\frac{9}{4}k_1-\frac{1}{12}m_3-\frac{9}{2}
\lb{a14}
\ee
We see that there are 9 non-vanishing coefficients. Obviously, not all of them can be  independent. Indeed, 
we find the following relations between these coefficients,
\be
&&h_{22}+3h_{21}=0\nonumber \\
&&h_{33}+27h_{31}+9h_{32}+18\eta=0\nonumber \\
&&h_{h44}+9h_{42}+27h_{41}+3h_{43}-\frac{9}{2}\eta=0
\lb{a15}
\ee
So that only 6 coefficients are independent. As these 6 independent coefficients we choose: $h_{21}$, $h_{31}$, $h_{33}$, $h_{41}$, $h_{42}$, $h_{44}$.
The resultant expression for $H(a,b,c)$ is given in the main text.

\section{Isotropic cosmology}
\setcounter{equation}0
\numberwithin{equation}{section}

Here we present a complete analysis of the isotropic case. The material in this appendix is not entirely original, as the model under consideration is the well-known Starobinsky model \cite{Starobinsky:1980te}. We include this discussion, however, to clarify some points that are not always made explicit.

 We consider a general time evolution with a scale factor $a(t)$,
\be
ds^2=-dt^2+a(t)^2(dx^2_1+dx^2_2+dx^2_3)\, .
\label{51}
\ee
For this metric the Weyl tensor vanishes, as is well-known. The Euler density on the other hand takes the form, 
\be
E_4=\frac{24\ddot{a}\dot{a}^2}{a^3}\, .
\ee
The  components of the quantum stress-energy tensor $T_{\mu\nu}$ then are represented as
\be
T^0_0=\hat{G}(t)\, , \  \   T^i_j=\hat{F}(t)\delta^i_j\, , \  \   i,\, j=1,2,3 
\lb{51-1}
\ee
The conservation equation $\nabla_\mu T^\mu_\nu=0$  produces only one non-trivial constraint,
\be
\partial_t \hat{G}(t)+\frac{3\dot{a}}{a}\hat{G}-\frac{3\dot{a}}{a}\hat{F}=0
\lb{51-2}
\ee
The further constraint comes from the anomaly, 
\be
\hat{G}(t)+3\hat{F}(t)=-{\cal A}(t)\, , \  \   {\cal A}(t)=mE_4=\frac{24m\ddot{a}\dot{a}^2}{a^3}\, ,
\lb{51-3}
\ee
where $m=\frac{GA}{2\pi}$, $A$ is the conformal charge. We assume that $m>0$. Substituting this into equation (\ref{51-2}),
one finds that
\be
\partial_t \hat{G}+\frac{4\dot{a}}{a}\hat{G}=-\frac{24m\ddot{a}\dot{a}^3}{a^4}\, .
\lb{51-4}
\ee
This differential equation can be solve as follows,
\be
\hat{G}(t)=-\frac{6m\dot{a}^4}{a^4}+\frac{C_0}{a^4}\, ,
\lb{51-5}
\ee
where the second term is a solution to the homogeneous equation, when the r.h.s. in (\ref{51-4}) vanishes, $C_0$ is an integration constant.
The freedom to choose  $C_0$ is due to the choice of the quantum state.	Non-vanishing $C_0$ can be interpreted as due to  the presence of isotropic radiation. 
One finds that,
\be
\hat{F}(t)=-\frac{C_0}{3a^4}+\frac{2m\dot{a}^4}{a^4}-\frac{8m\ddot{a}\dot{a}^2}{a^3}\, .
\lb{51-6}
\ee
The semiclassical gravitational equations take the form,
\be
G^\mu_\nu=T^\mu_\nu\, , \  \  G^\mu_\nu=R^\mu_\nu-\frac{1}{2}\delta^\mu_\nu R\, .
\lb{51-7}
\ee
For the isotropic metric (\ref{51}) one finds for the components of the Einstein tensor,
\be
G^0_0=-\frac{3}{\dot{a}^2}{a^2}\, , \  \   G^i_j=-(\frac{\dot{a}^2}{a^2}+\frac{2\ddot{a}}{a})\delta^i_j\, .
\lb{51-8}
\ee
The $(00)$ component of equation (\ref{51-7}) then leads to equation
\be
6m(\frac{\dot{a}}{a})^4-3(\frac{\dot{a}}{a})^2-\frac{C_0}{a^4}=0\, .
\lb{51-9}
\ee
It is solved as follows,
\be
{\dot{a}}^2=\frac{1}{4m}(a^2\pm \sqrt{a^4+\frac{8mC_0}{3}})\, .
\lb{51-10}
\ee
The behaviour  of the solution then depends on the value of the integration constant $C_0$.

\bigskip

\noindent {\bf 1. $C_0=0$.} In this case equation (\ref{51-10}) reduces to (excluding the possible solution when $a(t)={\rm const}$),
\be
\dot{a}^2=\frac{a^2}{2m}\, .
\lb{51-11}
\ee
The solution to this equation, $a(t)\sim e^{\pm \frac{t}{\sqrt{2m}}}$, is free of singularity and it describes de Sitter space-time. 
This is the solution found in \cite{Starobinsky:1980te}. 

\bigskip

\noindent {\bf 2. $C_0>0$.} In this case only plus sign can be taken in (\ref{51-10}). In the regime when $a(t)^4 \gg \frac{8mC_0}{3}$ the solution still describes the
exponentially growing de Sitter like Universe. On the other hand, for earlier times the solution approaches a singularity, where $a(t)$ vanishes. Assuming that this happens
for $t=0$ one finds that $a(t)\sim t$ in this case. 

\medskip

\noindent {\bf 3. $C_0<0$.} 
In this case, both the plus and minus signs in equation (\ref{51-10}) are allowed.
For the plus sign, the solution in the regime of large $a(t)$ still approaches a de Sitter phase, with $a(t)$ growing exponentially.
For the minus sign, by contrast, in the regime of a large scale factor the solution exhibits a power-law behaviour, $a(t) \sim \sqrt{t}$, as in a radiation-dominated universe.
In both cases, one has $a(t) \geq  a_m$, $a_m=(\frac{-8 m C_0}{3})^{1/4}$, so the scale factor never vanishes.
However, at the minimal value $a = a_m$, the first derivative $\dot{a}$ remains finite while the second derivative $\ddot{a}$ diverges.
Near this point (assumed to occur at $t = 0$), one finds to leading order in $t$ that
$a(t)=a_m(1+\frac{t}{2m^{1/2}}\pm \frac{\sqrt{2}}{6m^{3/4}}t^{3/2})$.
This leads to a curvature singularity at $t=0$. Note that the corresponding isotropic metric near the singularity is not of the Kasner type.

\medskip

We conclude that the isotropic case is singularity-free only if the integration constant $C_0$
 vanishes. This corresponds to a particular choice of the quantum state. This choice is, in a sense, distinguished by the fact that the resulting quantum stress-energy tensor can be expressed in terms of local curvature invariants, a representation that is possible when the metric is conformally flat (see \cite{Birrell:1982ix}) and was employed in \cite{Starobinsky:1980te}.

\end{document}